# Free electrons unveil nanolaser physics at the sub-picosecond and nanometer scale

*Cléo Santini[1,2], Thi-Huong Ngo[3] Luiz H. G. Tizei[4, 5], Aurélie Lloret[2], Tom Fraysse[1], Sébastien Weber[1], Adrien Teurtrie[1], Albert Polman[6], Virginie Brändli[3], Sébastien Chenot[3], Denis Lefebvre[3], Stéphane Vézian[3], Hugo Lourenço-Martins[1], Christelle Brimont[2], Benjamin Damilano[3], Thierry Guillet[2], and Sophie Meuret[1, *]*

[1]Univ. Toulouse, CNRS, CEMES, Toulouse, France

[2]L2C, Univ. Montpellier, Place Eugène Bataillon, Montpellier, 34095, France

[3]CRHEA, Univ Côte d'azur, rue Bernard Grégory, Valbonne,06560, France

[4]Laboratoire de Physique des Solides, Université Paris-Saclay, CNRS, Orsay, 91405, France

[5]Bruker AXS LLC, Kirkland, WA, 98034, USA

[6]Center for Nanophotonics, NWO Institute AMOLF, 1098 XG Amsterdam, The Netherlands

*Semiconducting nanowire lasers (NWLs) constitute nanoscale modular and tunable light sources, essential components for integrated optoelectronics. However, advancing NWL operation requires characterization of their near-field and dynamics at the nanometer scale, which is hampered by the diffraction limit of light. Here, we show a non-perturbative measurement of the absolute number of photons in a GaN/InGaN nanowire nanolaser cavity above the lasing threshold with sub-picosecond resolution and maps of the lasing mode spatial profile at the nanoscale. For this we use photon-induced near-field electron microscopy while simultaneously measuring the photons far-field, unraveling the key nanoscale and time-resolved characteristics of a NWL in operation. With this technique we observed lasing with a pump-power-dependent time offset of 0.5–1 ps that reflects carrier thermalization and spontaneous emission and a pulse duration of 500 to 700 fs. From the spatial maps of the lasing mode, we observe TM-polarized azimuthal whispering gallery modes along the nanowire circumference, as well as two distinct Fabry-Perot modes along the nanowire length. These spatial maps allow us to understand which of the multiple complex cavity modes available in the nanowire are responsible for lasing paving the way for their practical application into devices.*

Semiconductors nanowire lasers (NWLs) are a promising new class of nanometer scale light sources[1–4], due to their low lasing threshold[5,6], strong wavelength tunability[7,8], and ultrafast dynamics[9]. Because of their nanoscale size, the NWL lasing dynamics is complex, involving the interplay between nanowire geometry, spatially varying material gain, and modal structure of the cavity, all at subwavelength scale [7,8]. Therefore, understanding and controlling NWL requires ultrasensitive probing of the lasing spatial modes and temporal dynamics at deep-subwavelength and picosecond time scale. So far, this has only been addressed using far-field [9] or perturbative near-field probes [10,11]. However, the obtained information is insufficient to identify the lasing modal distribution, complicating the customization of NWL designs for their incorporation into practical devices.

An alternative method to probe the near-field optically excited nanostructures is photon-induced near-field electron microscopy (PINEM). This technique has been exploited for enhancing the electron microscopy contrast of single carbon nanotubes[12], to observe electron-light Rabi-oscillations,[13] or performed attosecond spectroscopy[14]. PINEM is a pump-probe technique where a focused pulsed electron beam probes the z-component, parallel to the electron trajectory, of the optical near-field induced by a pulsed laser on a nanostructure. When passing through the excited near-field of an object, the electron exchanges quanta of energy with the optical field, resulting in an electron energy spectrum with sidebands spaced by the probe photon energy $\hbar\omega$. The amplitudes of the side bands are uniquely determined by the strength of the optical near field at the electron probe position and the absolute number of photons $N_0$ in the cavity.

In this article, we use a unique combination of synchronized micro photoluminescence spectroscopy (μ-PL) and PINEM [12] in a custom-built ultrafast transmission electron microscope (UTEM) (see **Fig. 1**) to unveil the dynamics and spatial profile of NWL lasing modes. We demonstrate quantitative and non-perturbative measurements of the absolute number of photons $N_0(t)$ in specific modes of the NWL with sub-picosecond and nanometer resolution, i.e., accessing the mode evanescent field with subwavelength resolution. By varying the delay $\Delta t$ between the laser pump and the electron probe and by tracking the sideband amplitudes, we reconstruct the dynamical properties of the NWL near field.

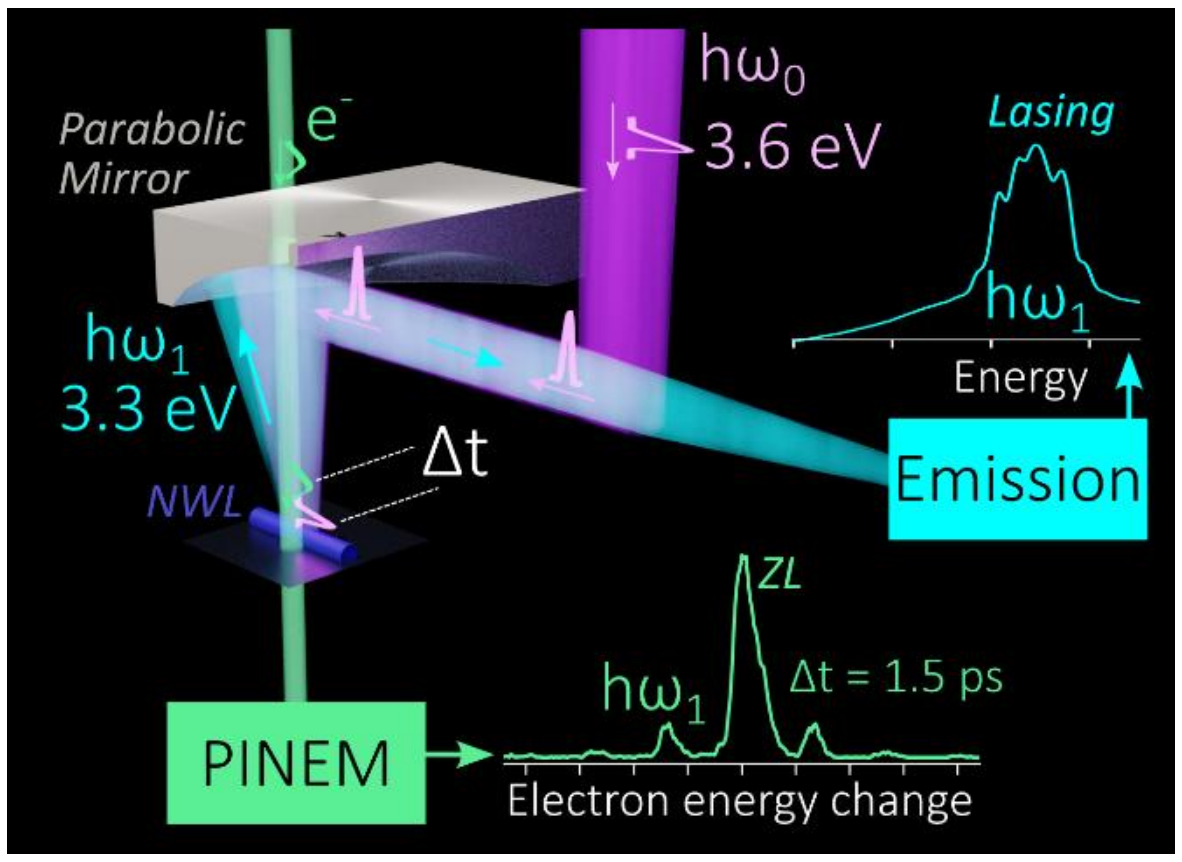

**Figure 1: Schematic of the experiment:** A pulsed laser beam (λ=316 nm, 280 fs) excites a nanowire laser (NWL) above the lasing threshold; a synchronized pulsed electron beam (200 keV, 400 fs) probes locally the NWL's near field at time $\Delta t$ after the laser excitation. The interaction strength is reflected in the sideband distribution in the electron energy spectrum on each side of the zero-loss peak (ZL). The emission spectrum of the NWL is simultaneously recorded to probe the far-field lasing spectrum.

**Results**

The UTEM used in this experiment is a modified Cold-FEG HF3300 Hitachi [15], in which a 250 fs ultrashort laser pulsed at 2 MHz is directed to both the electron gun (to generate a 400 fs pulsed electron beam) and to the sample via a parabolic mirror (see SI **Fig. S1**). The latter enables the efficient excitation of a single NWL with a laser spot size of approximately 10 μm, as well as the collection of its emission spectrum (μ-PL, see **Fig2-b)**. Crucially, the electron energy and the μ-PL spectra are recorded simultaneously, ensuring a correlated measurement of the near- and far-field of the NWLs. This is particularly important as the stimulated emission of the NWL is marked by the emergence of sharp PL peaks [6] at the cavity mode frequencies (**Fig. 2-b)**. The PL intensity versus input power of a typical NWL – both in the condition of illumination of the PINEM experiment and in a standard micro-photoluminescence (μ-PL) experiment – are shown in the Supporting Information (**Fig. S3** and **Fig. S10**).

### *Tracking the lasing cycle*

The nanolasers studied here are 5,5 μm long GaN nanowires with 40 nm InGaN inclusions every 180 nm (see **Fig. 2-a**), fabricated using a top-down approach (see SI) [16]. In order to follow the time evolution of the lasing dynamics of a single nanowire, we focused the electron beam at a fixed position on the side of the NWL in an aloof geometry and recorded electron energy loss and gain spectra for each delay $\Delta t$ between the pump laser pulse and the electron probe pulse, constructing a delay scan. Scans were recorded for a pump power below (P = 0.24 mW) and above (P = 0.66 mW) the lasing threshold (see **Fig. 2-b**). For negative delays $\Delta t < 0$ – corresponding to the electron probe impinging on the nanowire before the laser pump – the electron energy remained unchanged, and only the zero-loss peak (ZLP) was observed, as expected.

Under optical pumping, first below the nanowire lasing threshold (**Fig. 2-c**), at $\Delta t = 0$ distinct sidebands are observed at 3.6 eV, equal to pump photon energy, reflecting the PINEM interaction commonly observed from laser-excited dielectric nanostructures [17]. Then, for increasing time $\Delta t > 0$ each sideband shows a gradual spectral shift with time which we attribute to electron pulse energy-chirp [18] (0.89 ps/eV). This is due to Coulomb repulsion between electrons within the electron pulse [19,20] that increases the electron pulse width from 400 fs to 800 fs [18]. At any time $\Delta t > 0$, the energy spacing between each chirped sideband is constant at 3.6 eV, as expected in the chirp geometry.

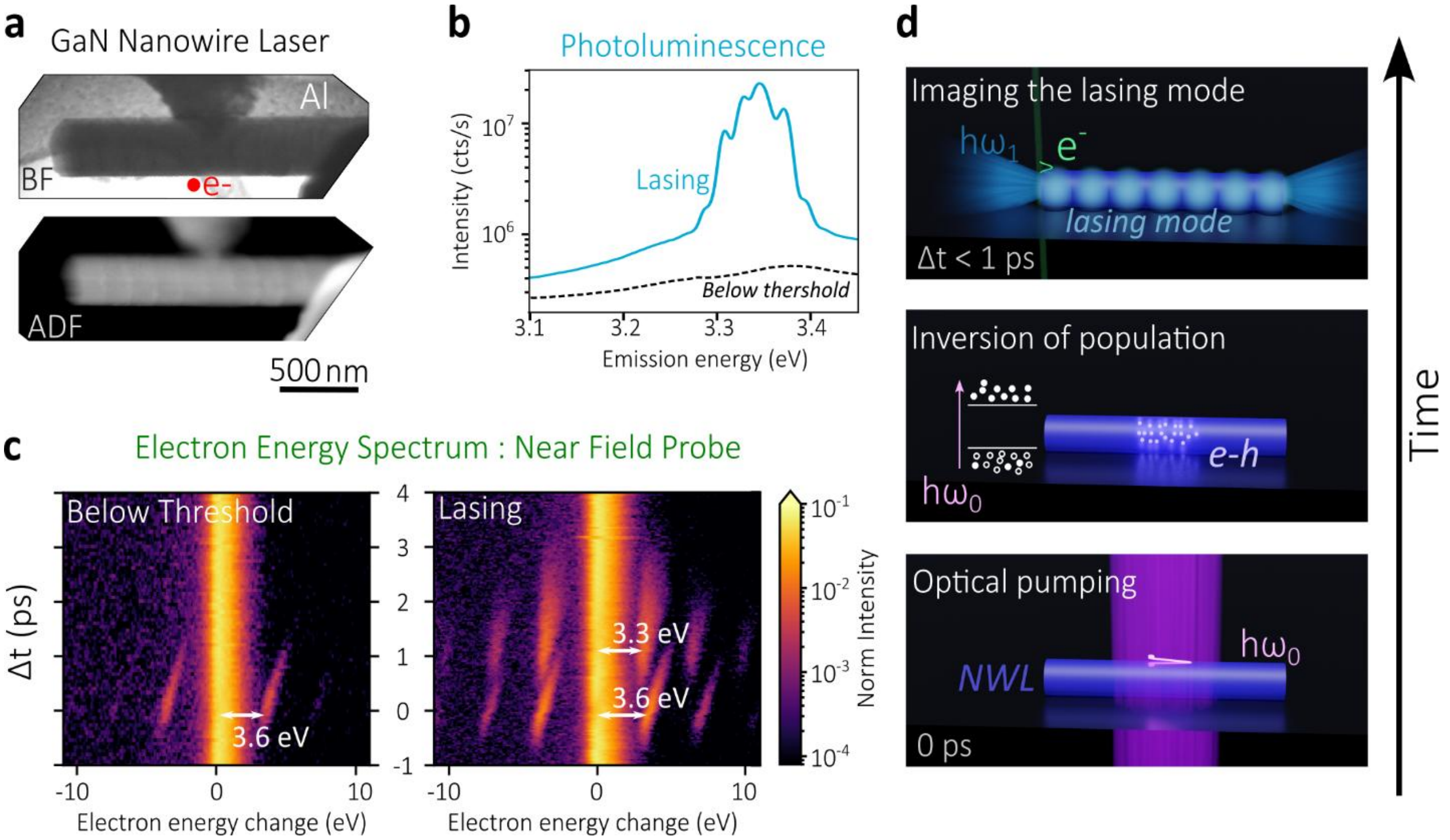

**Figure 2: Direct probing of stimulated electron-photon exchange with the nanowire lasing modes.** a) Bright Field (BF) and Annular Dark Field (ADF) scanning transmission electron microscopy (STEM) images of the nanowire under study. The position of the electron beam in an aloof geometry is illustrated on the BF image. b) Micro-photoluminescence (μ-PL) spectrum taken at the same time as the PINEM delay scan. Below the lasing threshold (pump power $P$=0.24 mW, dashed black line) and above the threshold ($P$=0.66 mW, blue line). The sharp peaks in the blue spectrum are characteristic of emission above the lasing threshold. c) Delay scan below and above the lasing threshold, showing the electron energy loss and gain around the zero-loss peak (0 eV) depending on the delay $\Delta t$ between the pump laser and the electron probe (200 keV, electron pulse 400 fs, laser energy 3.6 eV pulse duration 280 fs). d) Schematic representing the different steps of the lasing process from optical pumping to stimulated emission. The electron beam probes the NWL at a delay $\Delta t$ after the optical excitation of the nanolaser.

The same *direct scattering* phenomenon is observed when the NWL is pumped above threshold ($P > P_{th}$) **(Fig. 2-c)**. However, a second delayed set of sidebands also appears at $\Delta t > 0.5$ ps. These are separated by 3.3 eV, distinctly different from the pump laser energy. These bands correspond to the energy of the stimulated emission visible in the μ-PL spectrum (**Fig. 2-b**) and represent the interaction between the electron beam and the near-field of the stimulated emission in the NWL cavity. This represents the first an non-renormalizable second-order scattering process observed in PINEM, where the laser first creates intermediate electronic states in the sample; these states generate then, upon relaxation, cavity photons. These photons subsequently interact with the electron beam, thus resulting in an intrinsically incoherent, non-resonant and non-linear electron-photon interaction (See **Fig. 2-d**).

While several cavity modes are visible on the μ-PL spectra, the limited spectral resolution of our PINEM setup (~1.0 eV) prevents us from separating them in the electron energy spectra. This is primarily due to the energy dispersion of the electron source used in this experiment.

### ***Measuring the photon population with sub-picosecond resolution***

The amplitude of the n[th] sideband – corresponding to the electrons having exchanged $n\hbar\omega$ with the field – is noted $P_n$ with $n$ in $\in \mathbb{Z}$. Notably, all the amplitudes $P_n$ are uniquely determined by a single quantity - the electron-light coupling constant $g$ - which can be expressed as [21]:

$$g = \sqrt{N_0}\, g_0 \qquad 1$$

Here, $g_0$ denotes the electron-photon scattering amplitude that is determined by the spatial distribution of the near field and the electron velocity, a mode property independent of the laser excitation, which can be measured by electron energy-loss spectroscopy (EELS) [22]. Therefore, we can extract $g(t)$ from the delay scan (**Fig. 2-c**) and deduce the number of photons in the cavity $N_0(t)$ as a function of time.

As described above, we attribute the first set of electron energy sidebands ($\Delta t = 0$ ps) to the pump laser light directly scattered at the surface of the nanowire, and the second set of sidebands to the stimulated near field of the NWLs during lasing. We first extract the electron-light coupling constant $g(t)$ from the first set of sideband amplitudes $P_n$. Since these correspond to the interaction between an electron pulse (400 fs) that is longer than the light pulse (250 fs) the scattering can be considered incoherent (each electron of the wavepacket experiences a different field). In this case, the sideband amplitudes are modeled by a Poisson distribution [23]. The resulting fit of $g_{scat}(t)$ is shown in **Fig. 3-a** for three different pump powers above the lasing threshold. The width of these spectra is around 800 fs and matches with the convolution of the electron and laser pulse durations.

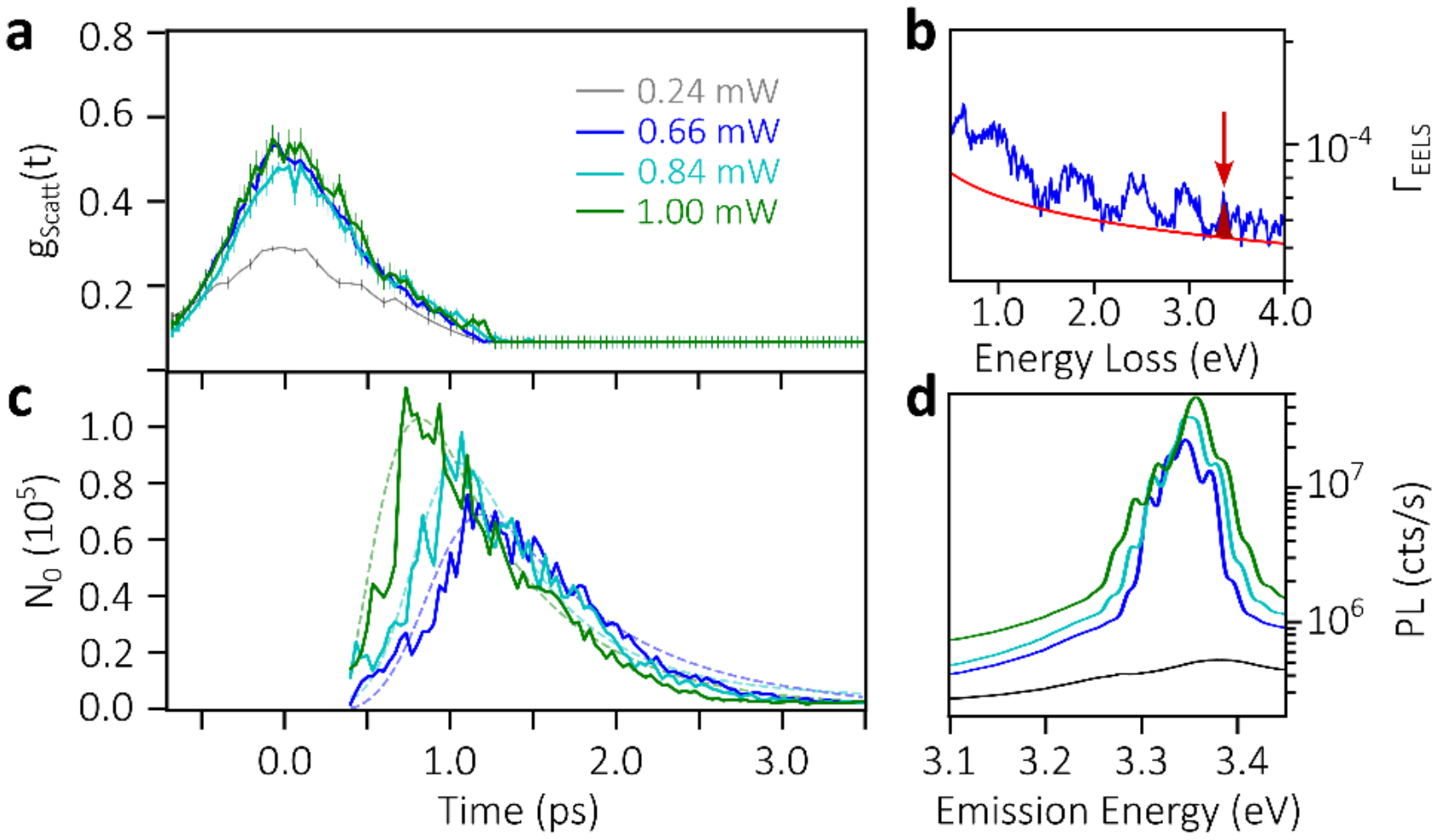

**Figure 3: Photon population in the cavity above the lasing threshold** a) The electron-near field coupling constant $g_{scat}(t)$ as a function of pump power at a fixed position. b) EELS spectrum acquired on the same position as the one where the delay scan of **Erreur ! Source du renvoi introuvable.** was acquired (same NWL). The peak intensity corresponding to the lasing mode ($E$ = 3.3 eV, red arrow) is fitted after removal of the background to extract $g_0$: the square root of the red area. c) Evolution of the number of photons in the cavity, $N_0(t)$ deduced from $g_{las}(t)$ in the lasing regime as a function of the pump power. d) NWL emission spectra taken simultaneously with the delay scans for the four pump-powers presented in a,c.

For interaction between the lasing mode and the electron beam, represented by the second set of sidebands, the optical near field persists longer than the duration of the electron wavepacket, resulting in a slight increase in the coherence of the interaction. The model used for the first interaction is therefore no longer applicable. The interaction was thus modeled as in [24] and the resulting $g_{las}(t)$ is extracted for the same three different pump powers (the detailed fitting procedure is explained in the SI). The variations of the coupling constant $g_{las}(t)$ directly reflect the photon population dynamics in the lasing cavity $N_0(t)$ (Eq. 1). To obtain a quantitative estimate of the photon population, we measured the electron-photon coupling constant $g_0$ of the same NWL using EELS on a 200 keV monochromated STEM (Nion HERMES by Bruker). The EELS spectrum taken at the same position as the PINEM delay scan (**Fig. 3-b**) shows a peak at 3.3 eV (marked by a red arrow), clearly visible just below the GaN band gap. Additional peaks are visible at lower energy corresponding to other cavity modes that are not coupled to the bath of spontaneous photons and therefore unavailable for lasing. From the EELS peak intensity at 3.3. eV we have estimated a value of $g_0 \sim 3.8\ 10^{-3}$ from which we then deduced the photon population dynamics $N_0(t)$ in the cavity for each pump power (**Fig. 3-c**). For a pump power of 1.0 mW a maximum of $10^5$ photons is reached 700 fs after excitation. As the PINEM interaction itself exchanges only 1 or 2 photons (gain or loss) the influence of the electron probe on the cavity modes and dynamics is negligible.

Figure 3-c shows a steep onset of the lasing mode density upon pumping, at a delay of 0.5-1.0 ps after the excitation pulse. We assign this to the carrier thermalization and radiative decay time need to fill the cavity with photons to the lasing threshold. To quantify this observation, the extracted $N_0(t)$ curves are fitted with an exponential decay convoluted with $g_{scat}(t)$, to take the electron chirp into account. From this analysis we find that the lasing onset time decreases from 0.85 ps at $P$ = 0.66 mW to 0.53 ps at $P$ = 1.0 mW, in agreement with the fact that population inversion is achieved earlier for higher pump powers [25,26]. From the fits we find that the NWL lasing pulse width is between 0.5 and 0.7 ps, in agreement with the ultrafast dynamics of nanowire laser modes[2,27].

**Nanoscale cavity mode imaging**

So far, we have discussed the temporal dynamics of the photons population inside the cavity; we will now turn to the spatial mapping of the near field of the lasing mode with sub-wavelength resolution. Due to their large diameters (between 300 and 500 nm), the NWLs support complex and multiple lasing mode profiles that can be broken down into two types of cavity excitations: Fabry-Perot modes (FPM) across the length of the nanowires and whispering gallery modes (WGM) across their circumference [26]. The far-field emission pattern is different for both type of modes (FB and WGM) as observed in μ-PL experiments (see SI **Fig S3**). However, this is an oversimplification, as some modes are complex combinations of both types. Measuring the near-field pattern of the lasing mode is critical to distinguish between these as they cannot be separated from the spectrum due to their spectral overlap and the numerous modes available at the same energy for this kind of large cylindrical cavity.

In order to characterize the near-field profiles of the bare cavity modes (i.e. in absence of external laser pumping), we performed STEM-EELS hyperspectral imaging on the same monochromated Nion HERMES machine - a technique which had already demonstrated its efficiency in probing such modes [28,29]. We identified clear peaks on the side of the NWL, which we assign to a WGM. As the electron beam only couples to field components along the trajectory, the imaged modes are TM polarized.

In contrast, a broad spectrum is observed when probing EELS on the top of the wire, corresponding to a wide range of FPM. For the latter the free spectral range (FSR), given the cavity length of a few um, is less than 30 meV which is below the EELS spectral resolution of the microscope at 200 keV, and so these form a broad background [28] (SI - **Fig. S13**). Fitting the EELS hyperspectral map around 3.3 eV we can separate the contribution of the WGM (amplitude of the peak) and of the FPM (amplitude of the continuous background). **Fig. 4-a** shows the different spatial variations of the cavity profile, for E = 3.3 eV, for FPM most intense at the end facet and for the WGM along the edge.

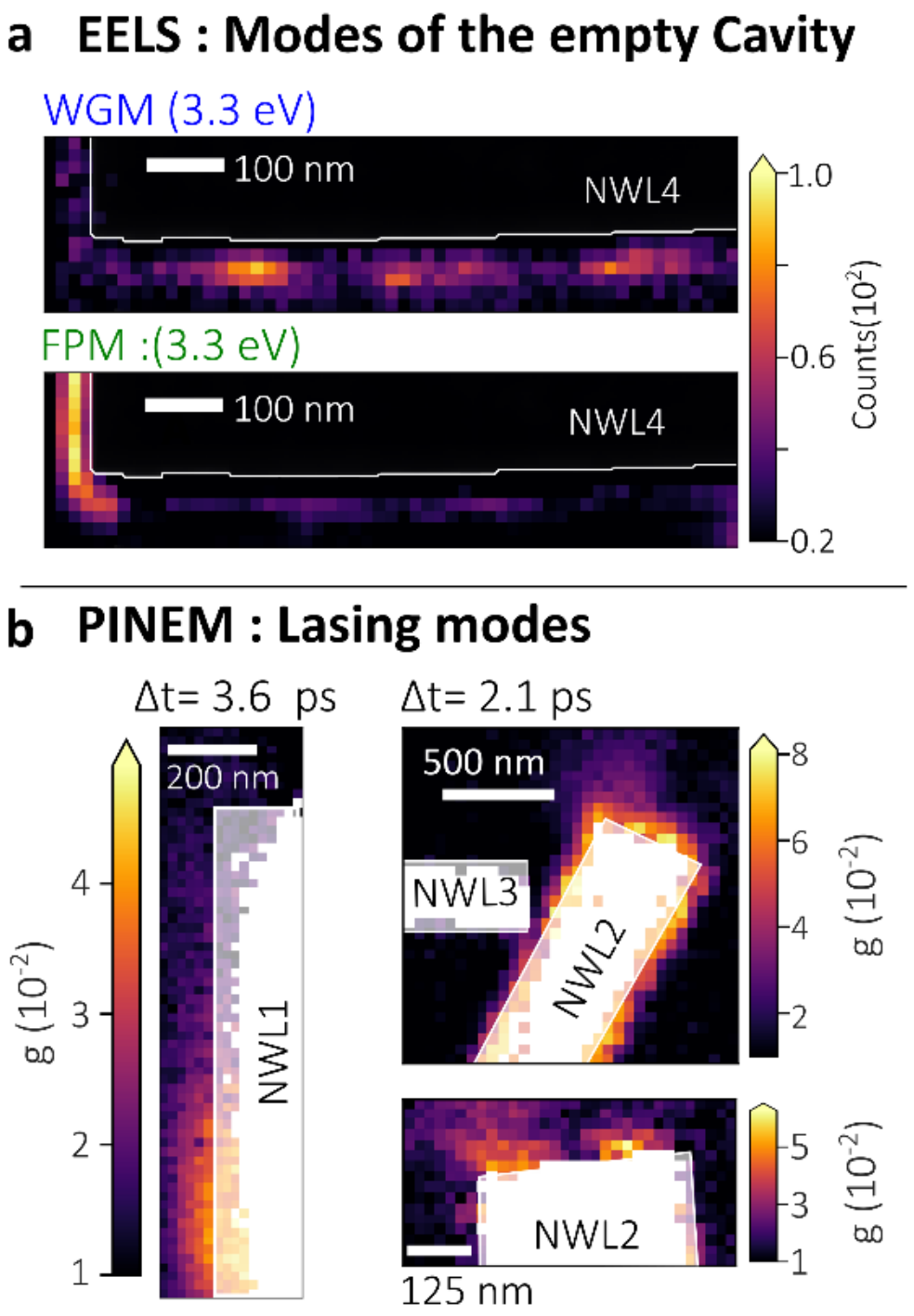

**Figure 4: Lasing mode spatial profiles.** a) Electron energy loss map (without laser pumping) at 200 keV using a monochromated Hermes Nion dedicated STEM on a typical NWL. Spectral contributions at 3.3 eV of whispering gallery mode (WGM, top) and Fabry-Perot mode (FP, bottom). b) PINEM on NWLs above the lasing threshold. NWL1: Map of coupling constant $g$ along the nanolaser at $\Delta t = 3.6$ ps and $P = 0.31$ mW (just above the lasing threshold). The near field of the lasing mode is localized on the side of the NWL corresponding to a WGM. NWL2: Maps of $g$ for $\Delta t = 2.1$ ps. Top: $P_{th} < P = 0.88$ mW, the electron coupling constant to the lasing mode peaks at the end facet, corresponding to a FP mode. Bottom: $P_{th} < P = 1.2$ mW; zoom-in showing details of lateral lobes for a different lasing FPM (associated PL spectra in **Fig. S10** of the SI).

We now image the near-field of the NWL when optically pumped by the fs-laser above the lasing threshold. To do so, we perform the same PINEM experiment as before albeit here, the pump-probe delay $\Delta t$ is fixed while the impact parameter $\vec{R}$ of the electron beam is raster-scanned over the nanowire. The resulting $g(\vec{R})$-map for $\Delta t = 3.6$ ps (lasing NWL cavity) - is shown on **Fig. 4-b**. A delay of more than 3 ps after optical excitation was chosen, because the pump power is just slightly above the lasing threshold ($P = 0.31$ mW) and therefore the onset time is longer than in **Fig. 3-c** (see SI **Fig. S12**). The map at $\Delta t = 0$ ps (i.e. direct scattering of the pump laser on the nanowire) is shown in the SI (**Fig S15**).

For $\Delta t = 3.6$ ps, the induced near field is localized around two InGaN inclusions (see **Fig. 2a**), spaced by a distance of 180 nm. By comparison with the EELS experiment (**Fig. 4-a**), we attribute this near-field profile to a WGM, with an ortho-radial polarization compatible with the $z$-coupling of the PINEM technique. The role of the InGaN inclusion on the observed mode profile requires further investigation, and is outside the scope of this paper.

Nevertheless, the dominant lasing modes can vary strongly from nanowire to nanowire. Figure 4-b shows lasing modes that display clear features of FP-type modes. Depending on pumping conditions, two separated FB type modes were excited associated to two different μ-PL spectra shown in **Fig. S10** of the SI**.** A double-lobed structure is seen in the zoomed-in image in **Fig. 4-b** that is not visible on the low-magnification scan of **Fig 4-b**. Both are FB-modes with an expected strong near-field at the end facet, the nanoscale pattern is different, with only one clearly revealing two lobes positioned at the end face. Interestingly, **Fig. 4-b** depicts a neighboring nanolaser (NWL3) that does not exhibit lasing showing that using PINEM, we can differentiate the behavior of two NWLs located in close proximity. Together these measurements show the strong sensitivity of the ultrafast PINEM modal imaging technique for the identification of different types of lasing modes in different nanowires under different pump conditions.

**Conclusion**

In conclusion, the spatial and temporal distribution of lasing modes in GaN NWL was mapped with a few-nanometer and sub-picosecond resolution using PINEM. This technique induces negligible additional optical losses in the cavity, allowing us to probe the state of the nanolaser without being affected by the measurement. Combined with monochromated EELS, this method revealed the time-dependent photon population $N_0(t)$ within the cavity - without *a priori* knowledge of the light collection and focusing power - and the influence of the pump power on the lasing onset. Moreover, we mapped and identified the cavity modes, namely FPM and WGM, which were involved in the lasing regime.

This work demonstrates the potential of PINEM to study semiconductors population inversion dynamics, light absorption and field generation. In combination with the analytical capability of conventional TEM, PINEM enables to directly link the material's atomic structure heterogeneity (e.g., defects, impurities, strain), the surrounding environment, the geometry, and the optical properties of the NWLs.

It should be noted that the spatial, temporal, and energy resolutions of the UTEM used in this study are respectively limited to 10 nm, 400 fs, and 1.0 eV. Therefore, we expect that the new generation of monochromated high-resolution instruments [30,31] will be able to resolve challenges related to mode hopping, cavity mode interference, and inter-mode dynamics; diving even deeper into the physics of complex nanolasers[32].

Finally, quantum optical experiments with free electrons are currently an intensely investigated topic [33,34], yet, finding bright sources of non-classical light for these experiments remains a bottleneck. With semiconductors playing a key role in quantum technology, we believe that our work, even if it operates in a semiclassical regime with $10^5$ photons in the cavity at threshold, is a first step in the direction of using semiconductors in free electron quantum optics.

**Supplementary Material Content**

PINEM experimental set-up
Process for the fabrication of GaN/InGaN Nanowires
Nanowire Laser imaging spectroscopy in conventional μ-PL
Finding the coupling constant g
Simulation of the beam chirp
Photoluminescence spectrum during PINEM experiment
NWL1 Delay scan at the power P =0.31 mW
Supplementary data EELS study and analysis
PINEM at $\Delta t = 0$ ps for NWLs of **Fig. 3**
Delay scan without electron chirp above the lasing threshold.

**Corresponding Author**

*sophie.meuret@cemes.fr

**Author Contributions**

C. Santini, with the assistance of S. Meuret, carried out the entire PINEM experiments and data analysis for this article. She was also responsible for building part of the PINEM setup required for the experiment. T.H. Ngo and B. Damilano created the samples with assistance from V. Brändl, S. Chenot, D. Lefebvre, and S. Vézian. L. Tizei carried out the EELS experiment and analyzed the data alongside S. Meuret. T. Fraysse and H. Lourenço-Martins assisted with the PINEM interpretation and data analysis. A. Polman gave feedback on the manuscript. S. Weber and A. Teurtrie are in charge of the UTEM experiment and helped build a portion of the setup. S. Weber developed the acquisition software PyMoDAQ used in the experiment. A. Lloret oversees the optical setup for NWL characterization. C. Brimont and T. Guillet carried out the optical characterization and supervised the work with S. Meuret.

**Funding Sources**

This work received support from the National Agency for Research under the program of future investment TEMPOS-CHROMATEM (ANR-10- EQPX-50). The authors acknowledge financial support from the CNRS-CEA “METSA” French network (FR CNRS 3507) on the platform LPS-STEM. The authors acknowledge financial support of the French Agence Nationale de la Recherche (ANR), under the grant agreement ANR-23-CE09-0018-LUTEM. The work at AMOLF is part of the research program of the Dutch Research Council NWO.

**Acknowledgement**

The authors thank Nika Van Nielen for discussions and preliminary experiments. We also thank Mathieu Kociak for fruitful discussions and all the STEM group for their patience when the microscope was blocked at 200 keV. S.M thanks Dominik Lungerich for fruitful discussions.

**Supplementary Information: Free electrons unveil nanolaser physics at the sub-picosecond and nanometer scale**

## I. PINEM experimental Set-up

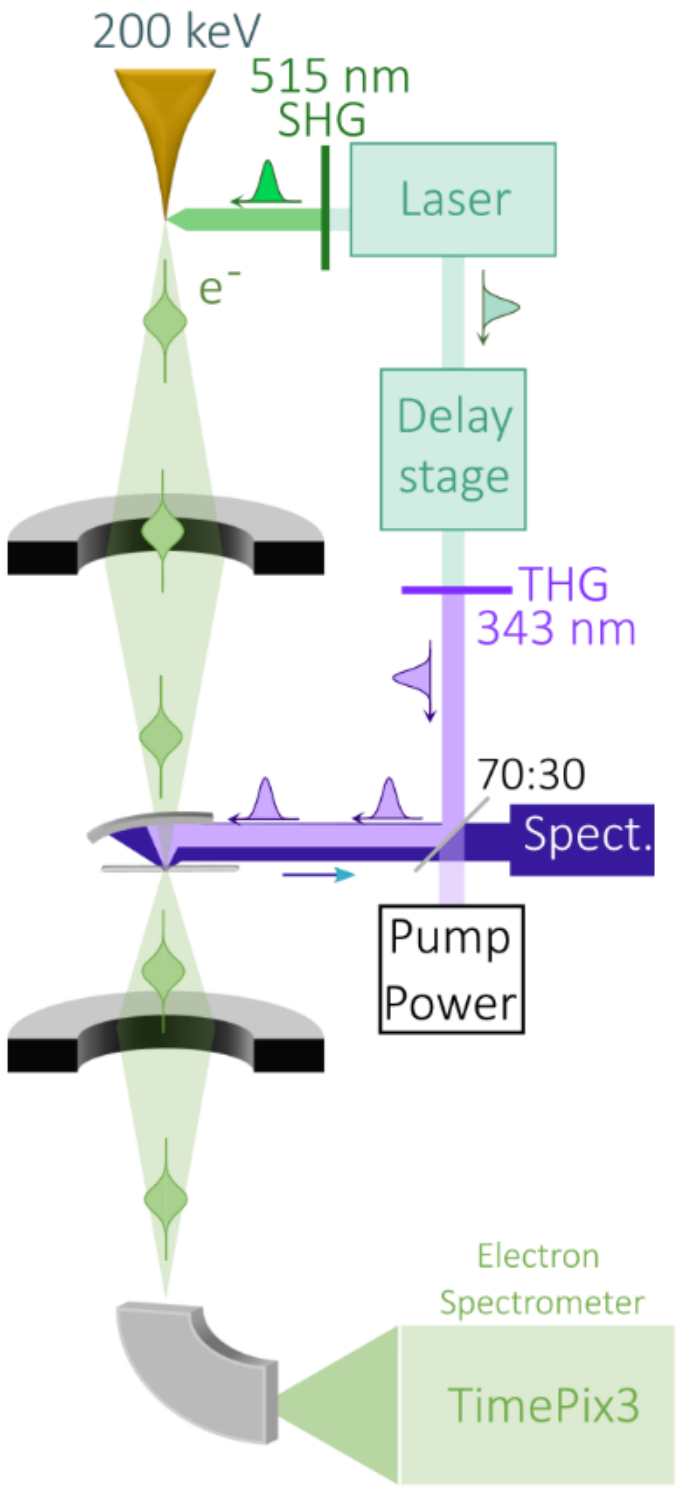

**TEM sample preparation:** TEM experiments require an electron-transparent substrate. The NWLs were deposited on a Holey carbon grid with an evaporated 100 nm Aluminium layer below the Carbon layer (the carbon layer playing the role of spacer between the Al layer and the NWL). This Al layer enables the determination of the pump-probe spatio-temporal overlap and seems to facilitate heat dissipation.

**UTEM set-up**: **Fig. S1** is a sketch of the experiment, where the Photoluminescence and PINEM spectrum are taken simultaneously thanks to a parabolic mirror that can efficiently inject and collect the light of the sample. The electron energy spectrum is recorded using a Gatan PEELS 666 spectrometer with an ASI Cheetah 3 timepix detector (**Fig. S1**)

**Fig. S1**: **Synchronous PINEM and µ-PL setup.** Illustration of the experiment: a 250 fs laser excites a Cold-FEG tungsten tip, resulting in a 2 MHz pulsed electron beam. The third harmonic (THG) of the same fs-laser optically excites the nanolaser via a parabolic mirror 200 µm above the sample. The photoluminescence is collected using the same parabolic mirror to track the nanolaser's emission. A motorized delay stage controls the delay between the optical pump and the pulse electron beam (probe). A spectrometer installed with a Timepix3 detector measures the scattered electron energy spectrum.

## II. Process for the fabrication of GaN/InGaN nanowires

The structure is first grown using metal-organic vapor-phase epitaxy on a sapphire substrate. It consists of a 3.3 µm-thick GaN layer and a multilayer with 10 periods of $In_{0.03}Ga_{0.97}N$ (40 nm) / GaN (180 nm). The fabrication process flow is shown in **Fig. S2**a-e. Starting with the epitaxial structure (**Fig. S2**a), resist is spin-coated onto the surface, and photolithography is used to create 1 µm-diameter holes with a 10 µm pitch (**Fig. S2**b). Then, 150 nm of Ni is deposited and lifted off to form the hard mask (**Fig. S2**c). The III-nitride layers are etched using $Cl_2$-based plasma in an ICP chamber (**Fig. S2**d). After etching, the Ni mask is stripped with piranha solution.

Subsequently, the nanowires are immersed in a 45 wt.% KOH solution at 100°C for 50 minutes to smooth their lateral facets and reduce their diameters (**Fig. S2**e). This process ultimately results in an array of GaN/InGaN nanowires with a height of 5.5 μm standing on the sapphire substrate, as depicted in **Fig. S2**f.

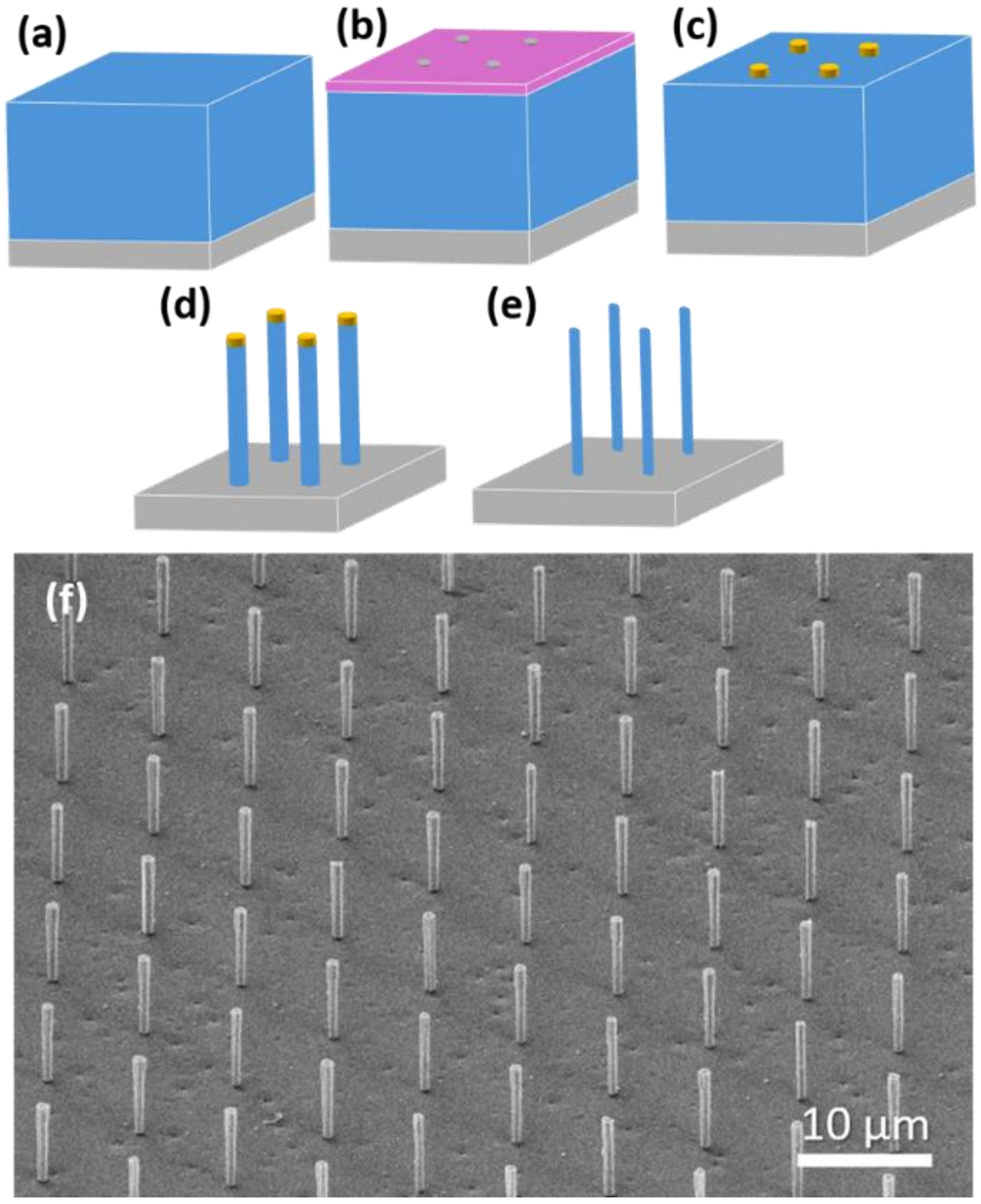

**Fig S2**: (a)-(e) Process flow diagram for nanowire fabrication: (a) growth by metal-organic vapor phase epitaxy on a sapphire substrate; (b) resist spin-coating and photolithography to create 1 μm diameter holes with a 10 μm pitch; (c) metallization and lift-off to form a Ni hard mask; (d) dry-etching using Cl2-based ICP; (e) Ni removal with piranha solution followed by KOH wet-etching to smooth the sidewalls and further reduce wire diameters. (f) 30°-tilted SEM images of nanowires standing on the sapphire substrate

## III. Nanowire laser imaging spectroscopy in a conventional μ-PL microscope

The nanowire samples are investigated in a standard micro-photoluminescence experiment, where lasing is observed under similar conditions as in the electron microscope: the nanowires are deposited on a membrane and optically pumped at room temperature with the second harmonics of a mode-locked TiSa laser, at 355 nm and pulse width of 150 fs. The pump is focused with a 50x microscope objective (spot diameter of the order of 1μm) and is partly absorbed by the nanowire. The emission is collected by the same objective and imaged onto the entrance slit of the spectrometer, so that the CCD detector provides a spectral and spatial image of the emission along the nanowire axis.

The spectra of the emission extracted at the end and the center of a given nanowire are represented in the **Fig. S3(a,b)** as a function of the pump power relative to the nanolaser threshold ($P_{th}$=34 pJ per pulse), and the corresponding images are shown in the panels (c-e). The emission spectrum below threshold is detected under the pump spot (panel c) and corresponds to the emission of the photo-generated electron-hole pairs in the InGaN sections of the nanowire, at 3.34 eV, and its LO-phonon replica at lower energy. At threshold, the lasing mode at 3.38 eV is also observed under the pump spot (panel d), whereas a second threshold corresponds to the appearance of a pair of lasing modes at 3.2 eV that are only detected at the two ends of the nanowire (panel e). Therefore, the lasing mode at 3.38 eV is attributed to a WGM, whereas the ones at 3.2 eV corresponds to Fabry-Perot modes. The power-dependent intensity of each kind of lasing mode is presented on the panel (f), showing the two successive thresholds and the nonlinear increase of the emission by a factor 10 to 40.

The lasing configuration depends on the investigated nanowire laser and on the size of the pump spot with respect to the nanowire length: in another nanowire, excited with a longer pump spot, the first lasing threshold corresponds to Fabry-Perot modes. These configurations are similar to the two PINEM observations.

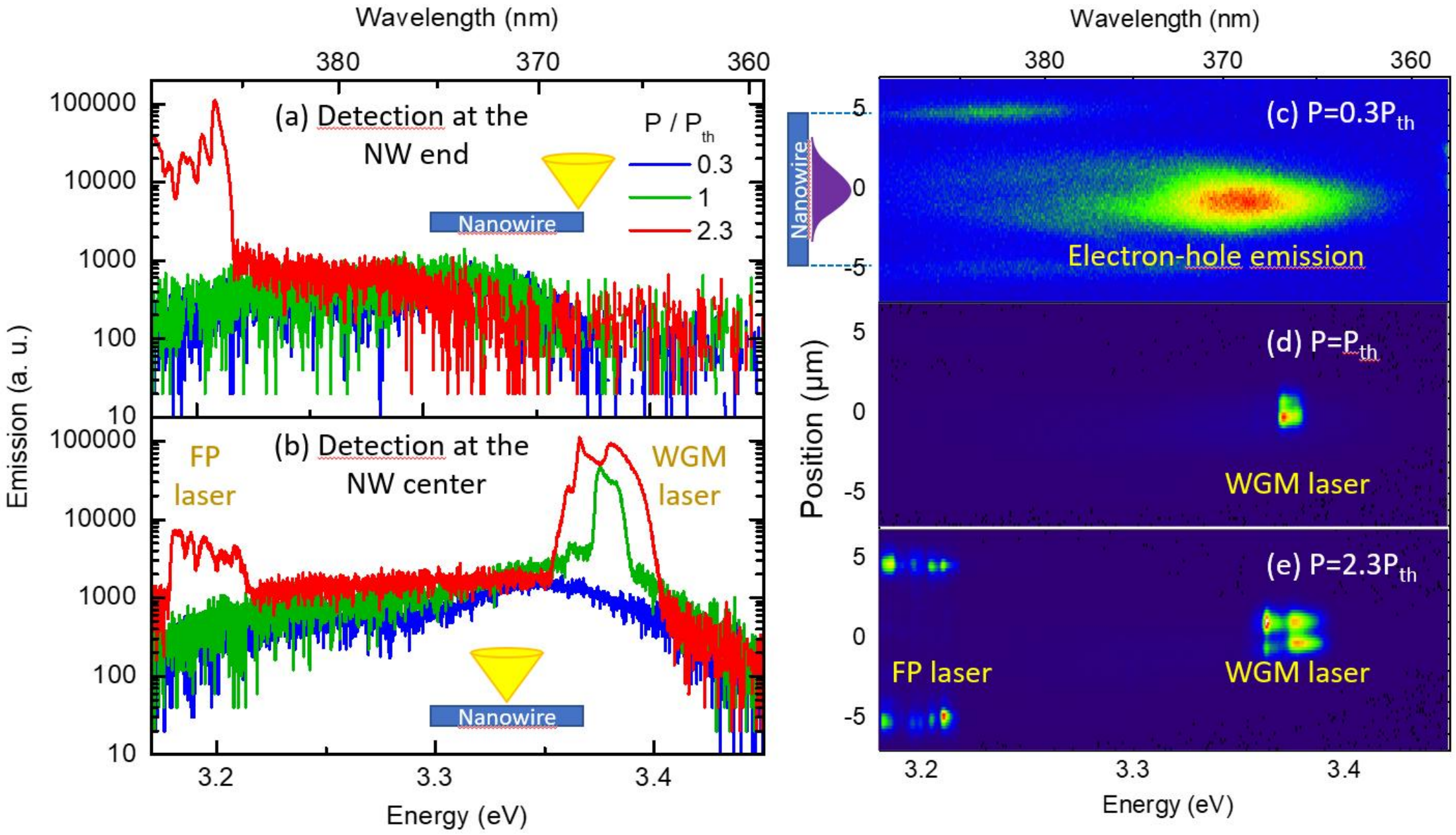

**Fig. S3**: Micro-Photoluminescence imaging of the nanowire emission; (a)-(b) Spectra recorded at the end and the middle of the nanowire vs pump power relative to the threshold $P_{th}$=34 pJ per pulse; (c)-(e) Spatially- and spectrally-resolved images of the nanowire emission, below, at and above threshold; (f) Intensity of the lasing modes vs pump power relative to the threshold

IV. Finding the coupling constant *g*

Everything will be illustrated with the Delay scan corresponding to an injection power of P = 0.66 mW and presented in **Fig. 2** of the main text.

**Raw Data treatment**

The two experimental data sets shown in **Fig. 2** of the main article are realigned to correct for the fluctuation of the zero-loss position on the spectrometer during dataset acquisition and normalized by the total intensity of the spectrum for each delay to account for photoemission process fluctuations. **Fig. S4** depicts the transformation of raw data into realigned data.

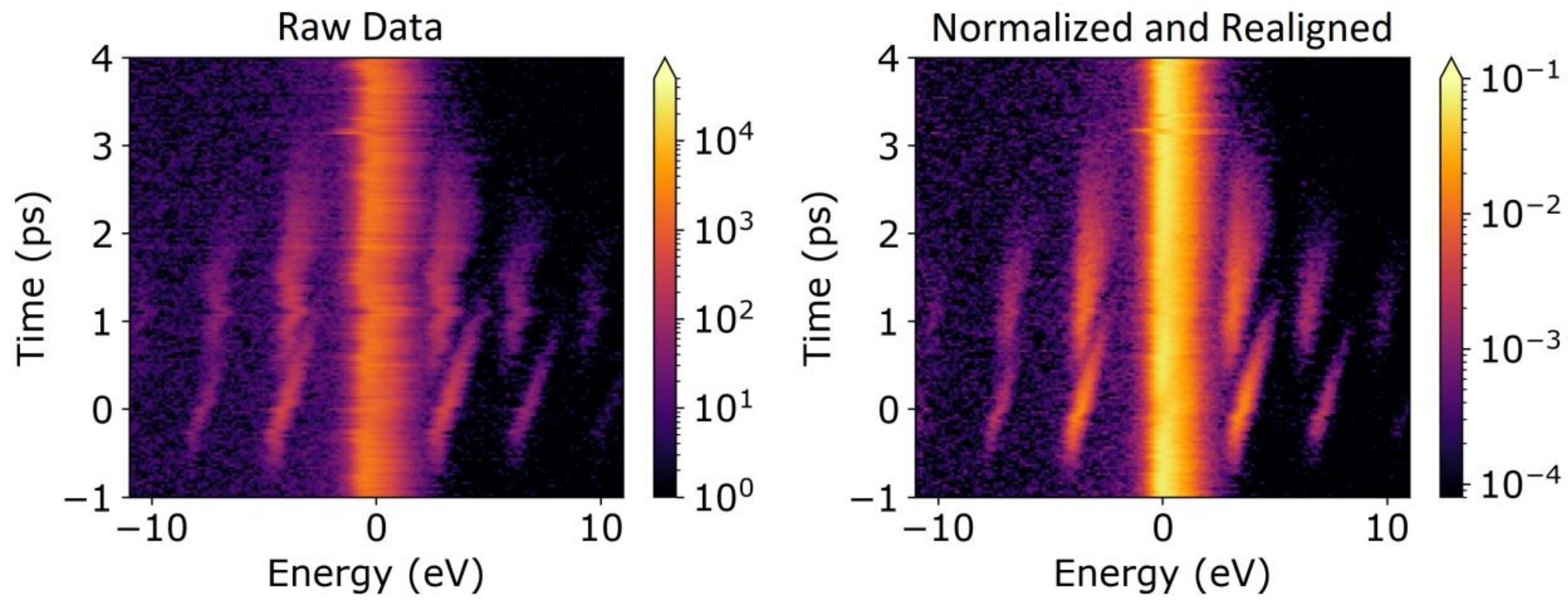

**Fig. S4**: **Raw Data to Realigned and Normalized Data**. Left: Raw Data for a pumping laser power of P=0.66 mW. Right: Processed data as shown on Figure 2 of the main text.

**Separating the two different contributions of the Delay scan**

Separating the two components allows us to fit the *g* from scattering and lasing independently. For this, we use a least squares method with a model with two sets of side bands, with the spacing between the side bands determined by $E_0$ and $E_1$. The amplitudes parameters were let free, but the loss and gain of the same $n^{th}$ order are forced to have the same amplitude. We obtain the two sets shown in **Fig. S5(a).** These two sets can be used to retrieve the two coupling constants $g_{scatt}$ and $g_{las}$, as shown in **Fig. 3** of the main text. One can see that the fit produces some residual side band at long delay; this is an artifact of the fit and accounts for the large error bar at long delay in **Fig. 3** of the main text.

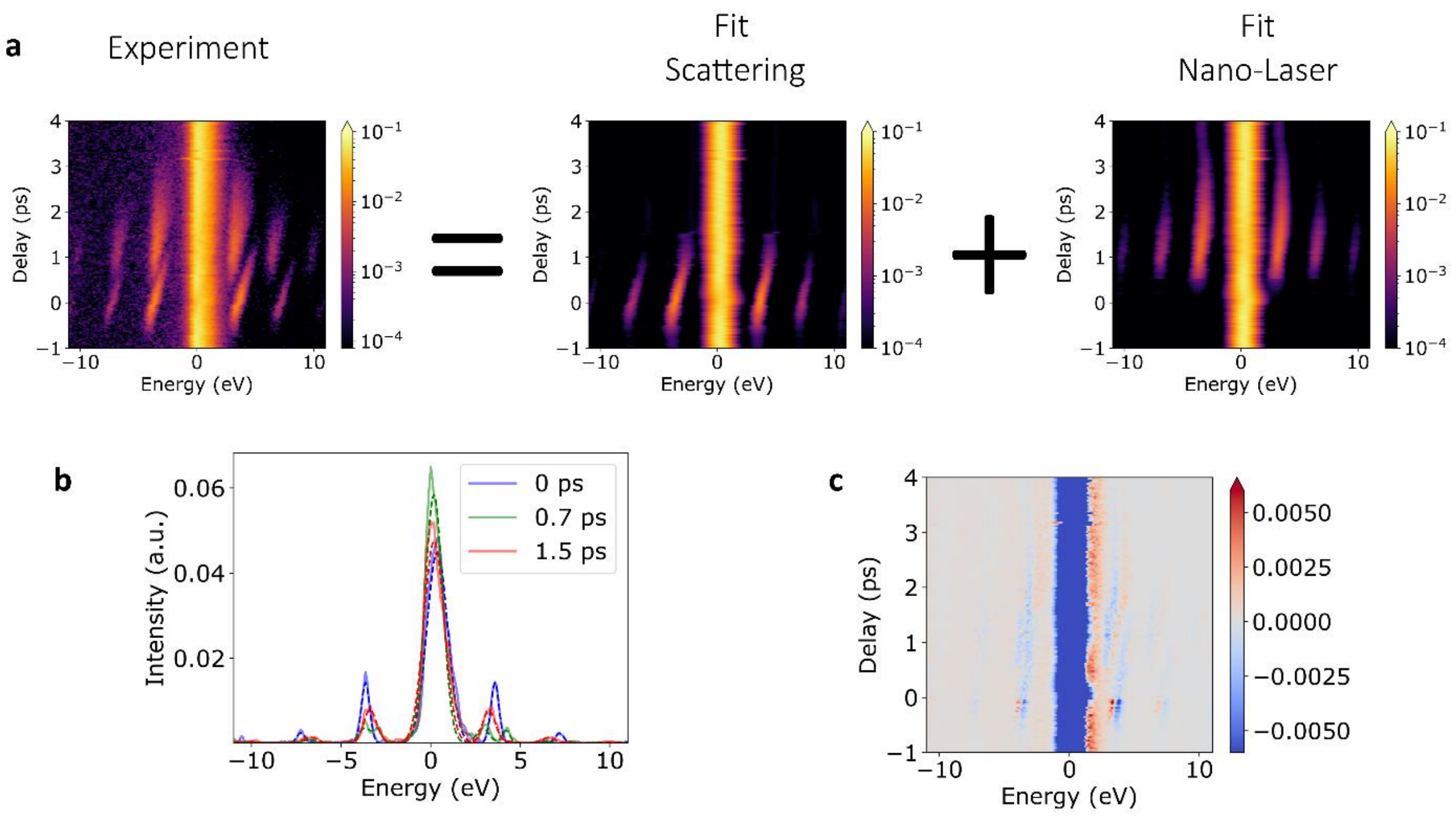

**Fig. S5**: a) The two components extracted from Raw Data with the fitting procedure b) Three extracted spectrum and their respected fits. c) Residual of the fit procedure.

## V. Fitting the two set of side band with PINEM theoretical model

Once the dataset is fitted with the model presented above, we fit separately the two sets of PINEM interaction found in **Fig. S5**. The scattering interaction is very nicely reproduced using the model of the weak interaction regime developed in [32], where the amplitude $A_n$ of the order n is given by :

$$A_n = e^{-2|g|^2} I_n(2|g|^2)$$

Where $g$ is the coupling constant, and $I_n$ is the modified Bessel function. Because of the electron chirp, calculating the amplitude of zero loss (n=0) is extremely difficult. As previously stated, we fixed the amplitude of the loss and gain to be identical (see above), so we fit the amplitude for n > 0 using the peak on the gain side.

The weak coupling regime cannot reproduce the first-to-second order ratio when interacting with the lasing mode. This could be explained by the longer-lasting field, as well as the difference in lasing field strength between pulses. It is entirely possible that population inversion and stimulated emission do not occur for every pulse. We apply the same model as in [21].

We calculate the value of *g* using these two fits (Figure 3 of the main article). **Fig. S6** depicts the delay scan not shown in the main text, from which the g for the other two pump powers were extracted (P=0.84 mW and P=1.0 mW).

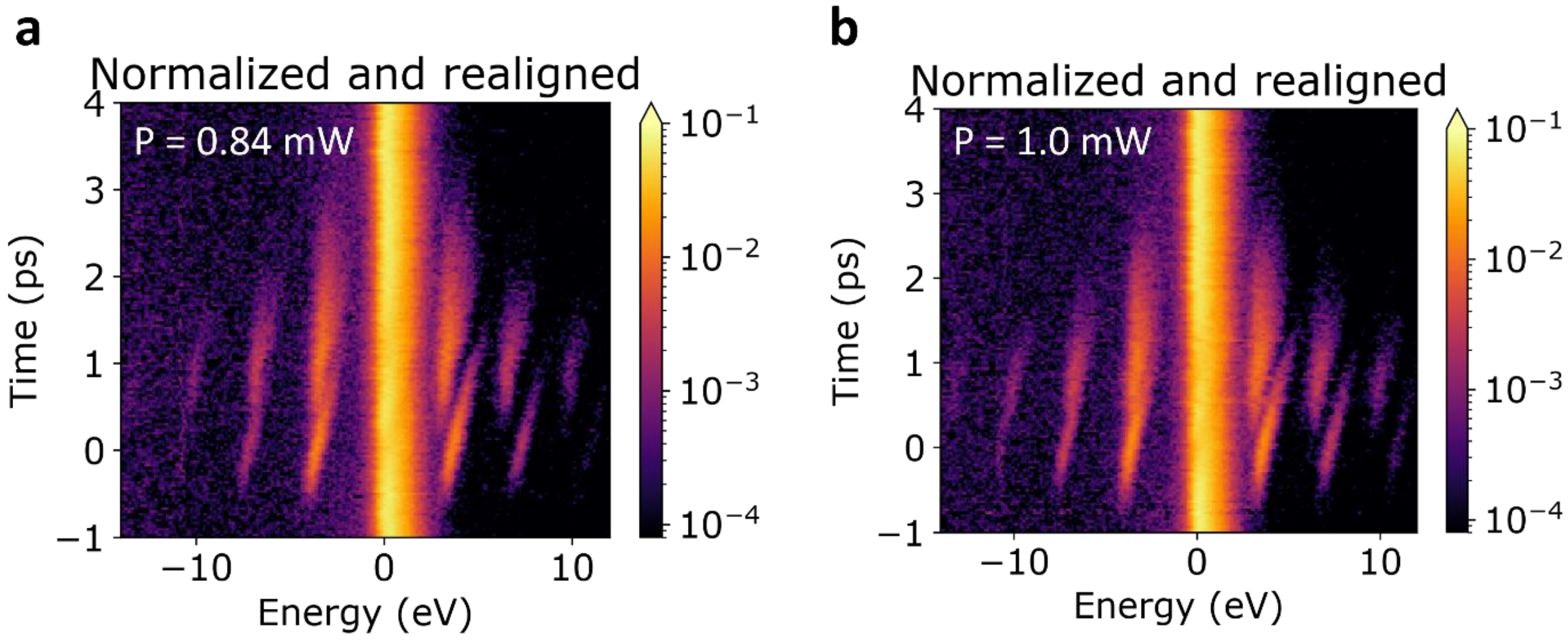

**Fig. S6**: Delay scan treated with the procedure explained in the section above for injection power P = 0.84 mW and P = 1.0 mW

## VI. Simulation of the beam Chirp

To simulate the chirp observed on the delay scan of **Fig. 2-C**, we convolute the chirp of the electron beam with the PINEM expected using the two models discussed in the previous.

To simulate the PINEM with no chirp we described the evolution in time of the coupling constant *g* (see **Fig. S7a**) by:

$$g_{scatt}(t) = \frac{A}{\sqrt{2\pi}\sigma_{pump}} e^{-\frac{t^2}{2\sigma_{pump}^2}}$$

$$g_{las}(t) = \frac{1}{\tau} e^{-\frac{t-t_0}{\tau}}$$

$$g(t) = g_{scatt}(t) + g_{las}(t)$$

We found the value of the onset time $t_0 = 0.85$ ps the characteristic time of the NWLs lasing pulse $\tau = 0.75$ ps by fitting the *g* curve of **Fig. 3** of the main article. We then use the two interaction models discussed in the section above, to reproduce the energy spectrum expected at every delay t. The result is shown on **Fig. S7-b**.

The electronic chirp P(t,E) is defined by [15]:

$$P(t,E) = \left(1/(2\pi\sigma_t\sigma_E)\right) * \exp[-(t - s_1 E)^2/(2\sigma_t^2)] * \exp[-E^2/(2\sigma_E^2)]$$

Where $\sigma_t$ and $\sigma_E$ are deduced respectively from the laser pulse width 282 fs and the energy spread of the source ΔE = 700 meV [13]. We extracted, from the slot of the first side band of the scattering interaction on the experimental data, a chirp of $s_1$ = 0.89 ps/eV). We then convolute the two resulting 2D map (depicted in Fig. **S7**) to obtain the simulated delay scan of **Fig. S8**

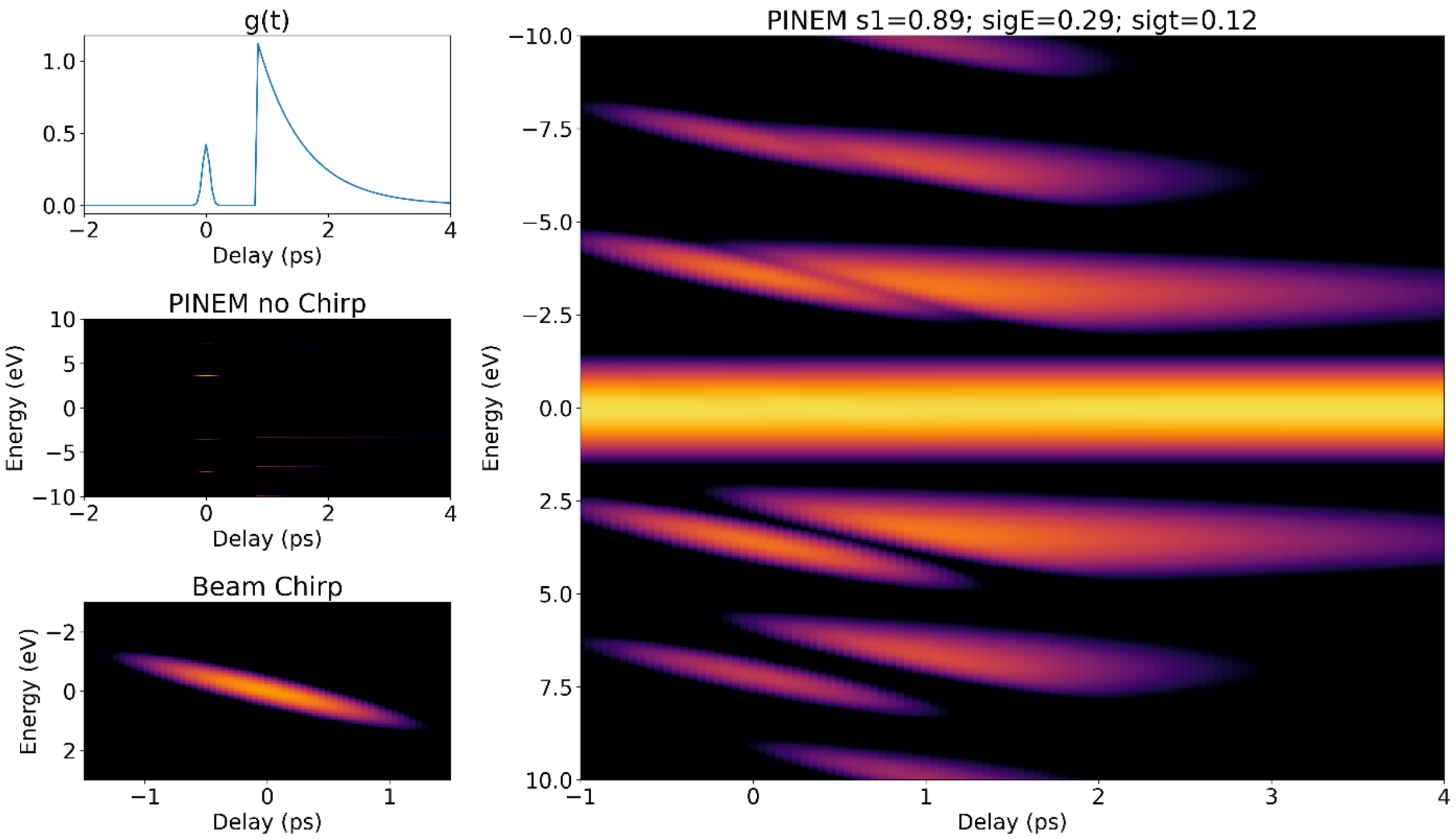

**Fig. S7**: Different steps of the chirp simulation. a) g(t) distribution without chirp, which correspond to the deconvoluted signal of the P=0.66 mW of Figure 1 of the main text. b) PINEM expected from the two models explain previously and with the value g(t) of a). c) Representation of the chirp from equation \ref{eq_chirp}. d) Convolution of the beam chirp of c) and the PINEM of b).

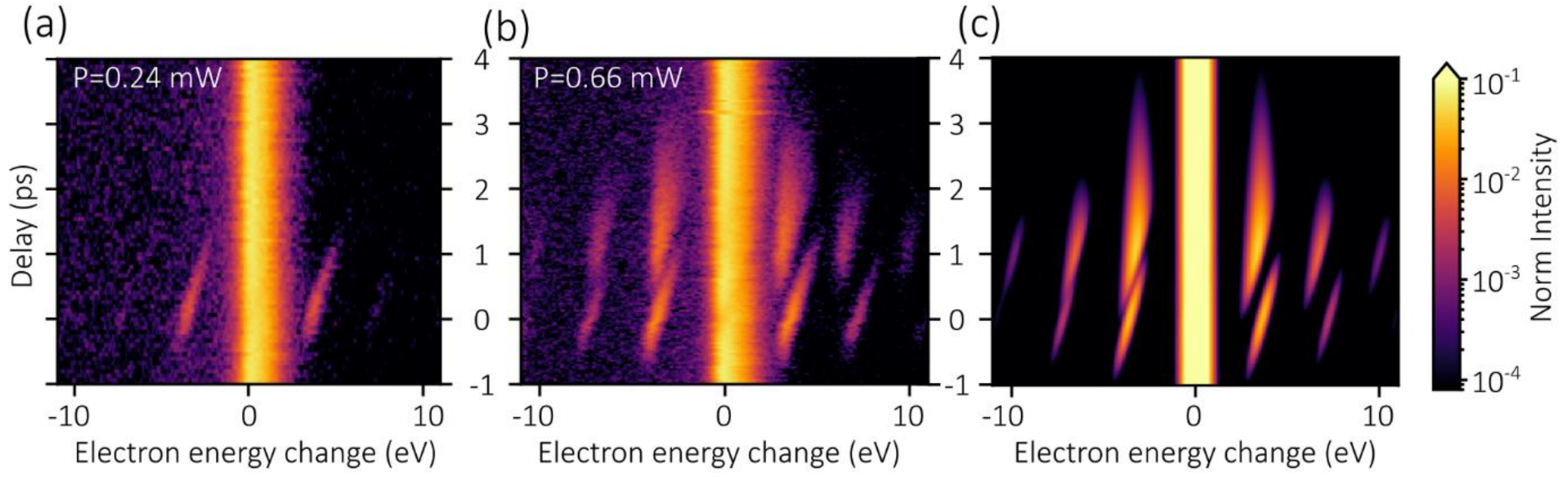

**Fig. S8**: PINEM of the stimulated near-field of a NWL. a), b) The two experimental datasets collected at a fixed position close to the nanolaser for pumping power levels below (P=0.24 mW) and above (P=0.66 mW) the lasing threshold. An electron energy spectrum is recorded every 33 fs delay between the laser pump and the electron probe. The data is normalized using the total intensity for each measurement. The noise visible only in the loss part of the spectrum is consistent with spontaneous electron losses standard in electron energy loss experiment. c) Simulation of b) taking with an onset time of 0.85 ps and a decay time of 0.75 ps for the laser mode and an electron chirp of 0.89 ps/eV.

## VII. Photoluminescence Spectrum during PINEM experiment

One PINEM map or delay scan is about 15 min, therefore the damage of a NWL under optical pumping as to be taken into account when performing multiple measurement of the same NWL. As explained in the main text the electron energy spectrum is recorded simultaneously with the photoluminescence, which allows us to track the evolution of the lasing emission properties for the duration of the scan. **Fig. S9** shows the luminescence spectrum of the four delay scans represented in **Fig. 2** and **Fig. 3** of the main text.

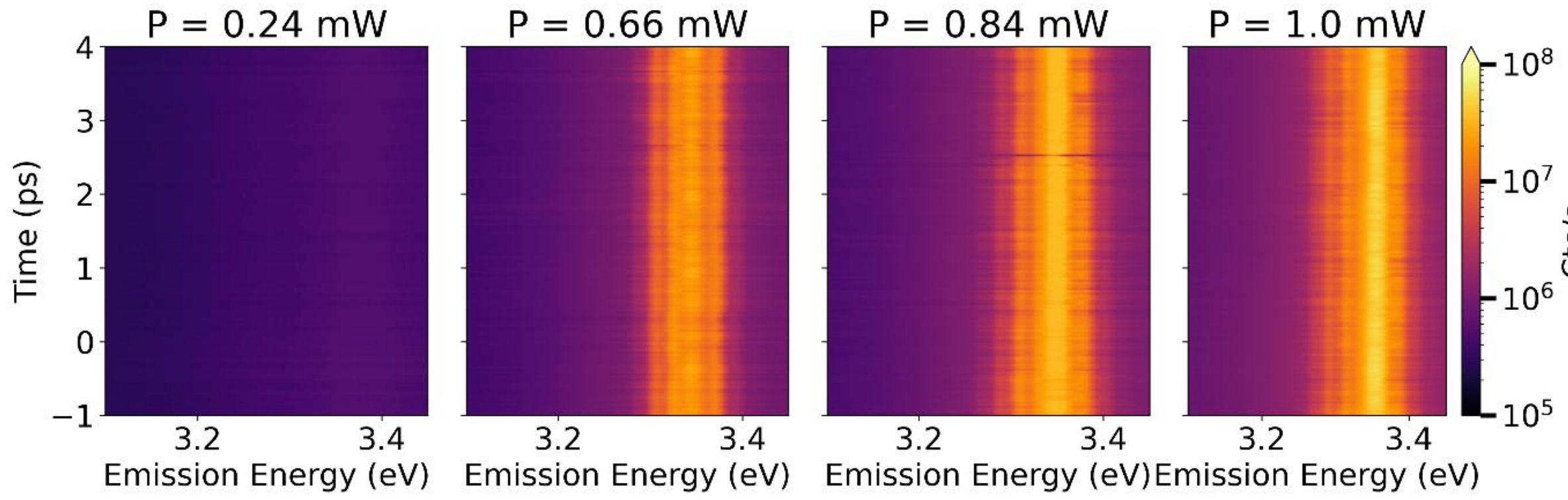

**Fig. S9**: Photoluminescence spectrum recorded for every step of the delay scan used for Figure 2 and 3 of the main text. We can see that for the injection powers 0.66 mW, 0.84 mW and 1.0 mW the NWLs is lasing at 3.3 eV and the emitted mode are stable for the all duration of the measurement. While for the injection power of 0.24 mW the NWLs is below the lasing threshold for the all duration of the delay scan. The emission spectrum depicted Figure 1 and Figure 3 are the sum over all the time step of the experiment.

In the same fashion, for the PINEM maps recorded at a fixed delay between the pump laser and the electron beam, the PL spectrum was recorded for every pixels the average spectrum is shown in **Fig. S10** and the PINEM delay scan on **Fig. S12.**

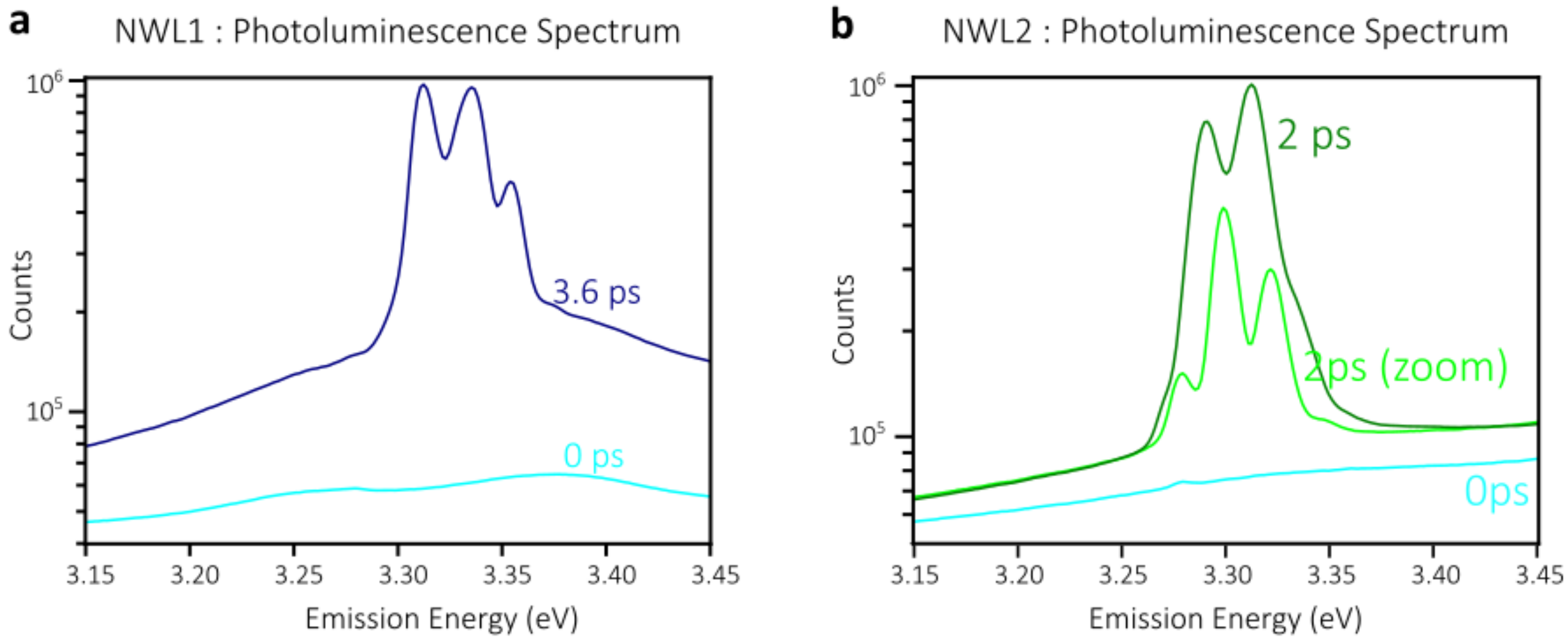

**Fig. S10**: **Average Photoluminescence Spectrum data of Fig. 4**. a) Photoluminescence of NWL1 for the map recorded respectively at 0 ps (P = 0.34 mW) and 3.6 ps (P = 0.31 mW). b) Photoluminescence of NWL2 for the map recorded respectively at 0 ps (P = 0.94 mW), 2.1 ps (P = 0.88 mW) and for the zoom around the top at 2.1 ps (P = 1.2 mW)

We note that the power used for the map at $\Delta t = 0$ ps (Fig. **S15**) is higher than the power used for the first map at $\Delta t = 2.1$ ps (Figure 4 of the main text). We recorded the PINEM map at $\Delta t = 0$ ps, once the NWL stopped lasing due to damage. Indeed, the map at 0 ps does not depend on the lasing properties but only on the scattering of the pump laser on the structure (see **Fig. 2** of the main text). We can also see that despite that both are FP modes, as shown on the map of **Fig 4**, the modes lasing for the two maps at 2.1 ps are not the same.

**Fig. S11** shows the intensity vs power within the same condition of illumination within the UTEM on another NWL from the same TEM grid. The characteristic S-shape expected for a lasing transition is recorded. The spectral linewidth reduces from a 100 meV below threshold to 8 meV (the spectrometer resolution is 3 meV) above threshold.

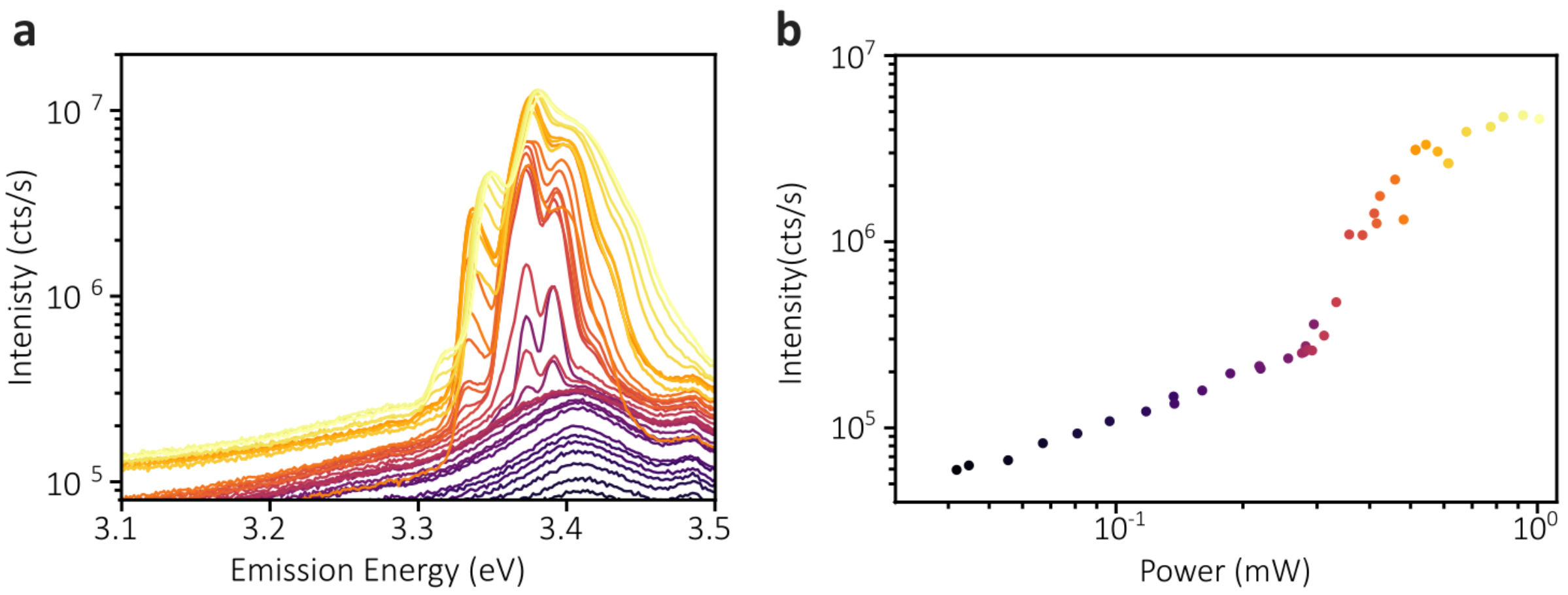

**Fig. S11**: a) Photoluminescence Spectrum of a NWL depending on the laser power. b) Intensity between 3.3 eV and 3.49 eV depending on power.

## VIII. NWL1 Delay Scan at the Power P = 0.31 mW

To avoid damaging the NWL1 during a relatively long acquisition we record the PINEM map of **Fig. 4** just above the lasing threshold at P = 0.31 mW. The delay scan is taken just before the PINEM map to find the right delay (for **Fig. 4** NWL1: Δt = 3.6 ps).

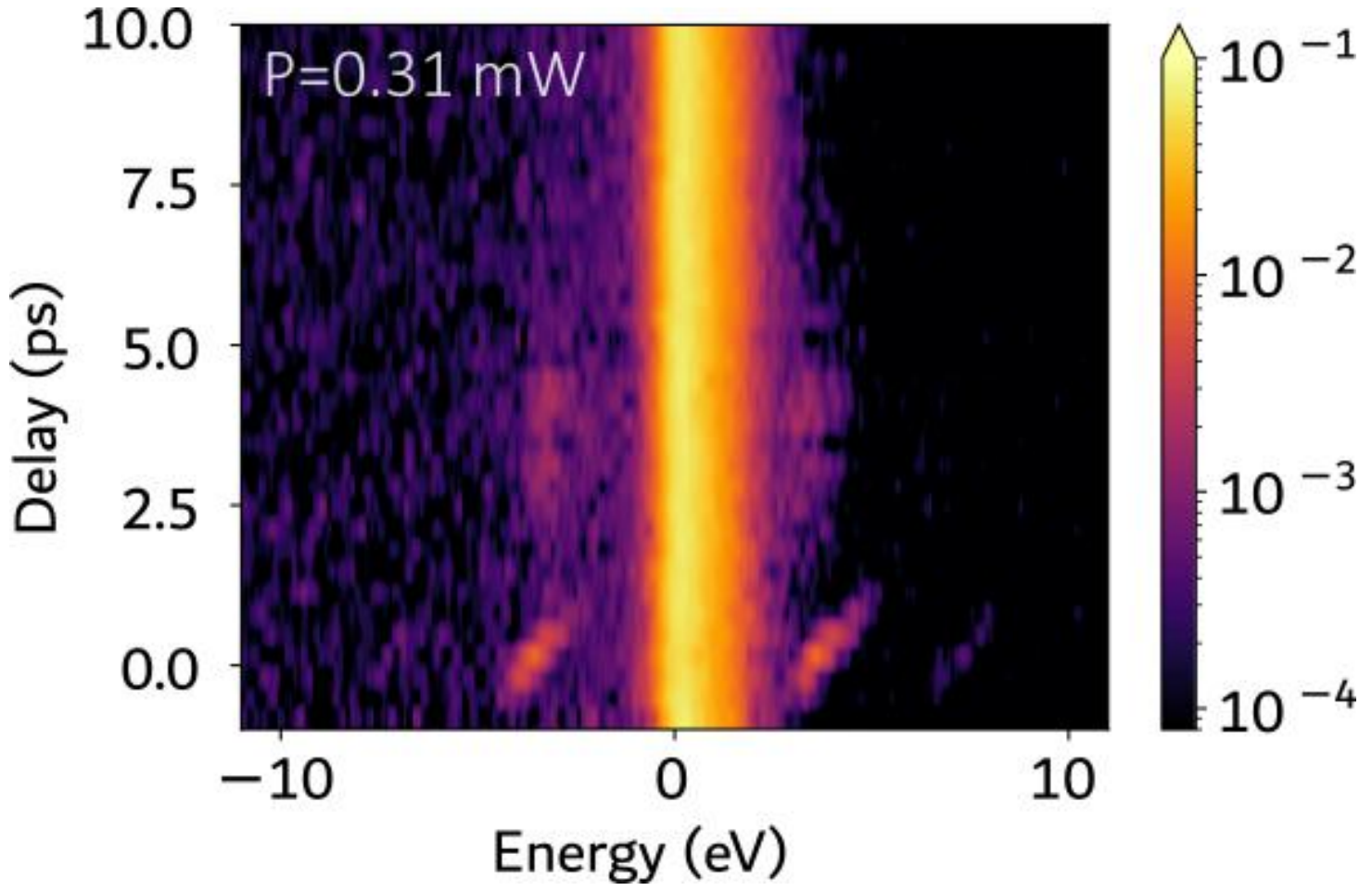

**Fig. S12**: Delay scan NWL1 at the same pumping power (P = 0.31 mW) as the map of Figure 4 of the main text.

## IX. Supplementary Data EELS study and data analysis

The EELS maps shown in **Fig. 4-A** of the main text are fitted maps from a hyperspectral dataset of NWL4.

**Fig. S13** shows the projection of the spectrum along the growth direction as well as the projection of a second set of data along the perpendicular direction showing the top of the nanowire. We clearly see the two behaviors explained in the main text.

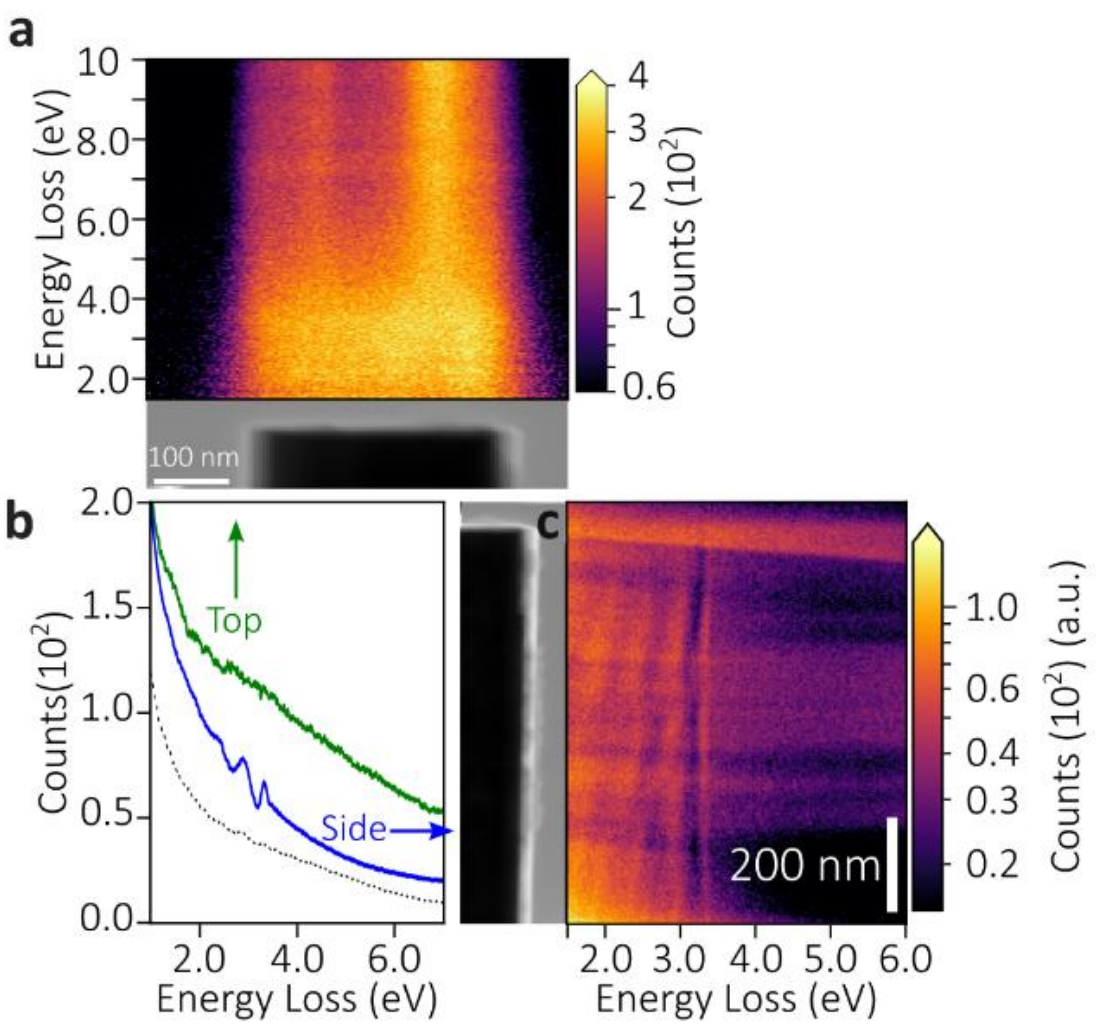

**Fig. S13** : EELS Spectral projections of a- the top of the nanowire and b-along the growth direction. The WGM are clearly visible along the side of the NWL while the top show what it seems to be a continuum of FPM.

**EELS Data from NWL 1**

The EELS spectrum presented in **Fig. 3** of the main text is extracted from a hyperspectral dataset taken on NWL1. The EELS spectrum of **Fig. 3** is taken from the same position than the PINEM delay scan. The rest of the hyperspectral data are summarized on the

**Fig. S13**. We can see the whispering gallery mode visible on the EELS spectrum of **Fig. 3** visible along the NWL on the projection map of

**Fig. S13c**.

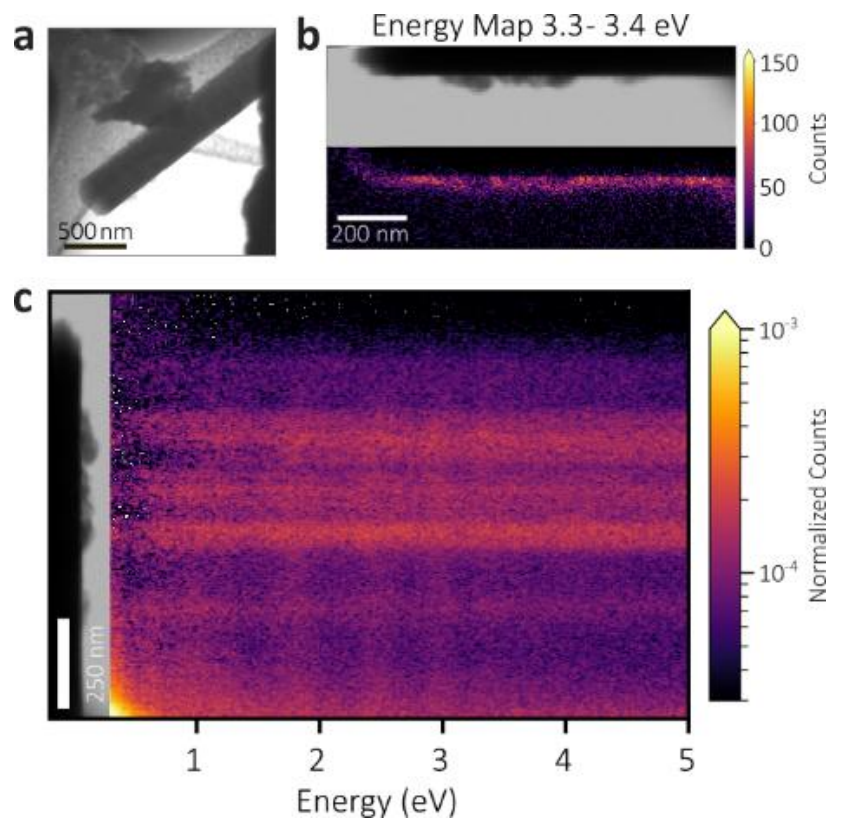

**Fig. S14**: **Monochromated EELS hyperspectral study of NWL1**. a) Bright Field (BF) Image Low Magnification b) Energy Map extracted from the data set from 3.3 to 3.4 eV with the corresponding BF image. c) Projection of the EELS Dataset along the NWL1 and after normalization and subtraction of the background.

## X. PINEM map at Δt = 0 ps for NWLs of Figure 4

For delay Δt = 0 ps (direct scattering), the intensity of the near-field depends on the geometry of the different interfaces, on the aluminum layer below the nanolaser as well as on the angle between the incident light and the sample due to the parabolic mirror [33]. The map of **Fig. S15** clearly show that the near-field pattern is drastically different between the two delays for each NWLs.

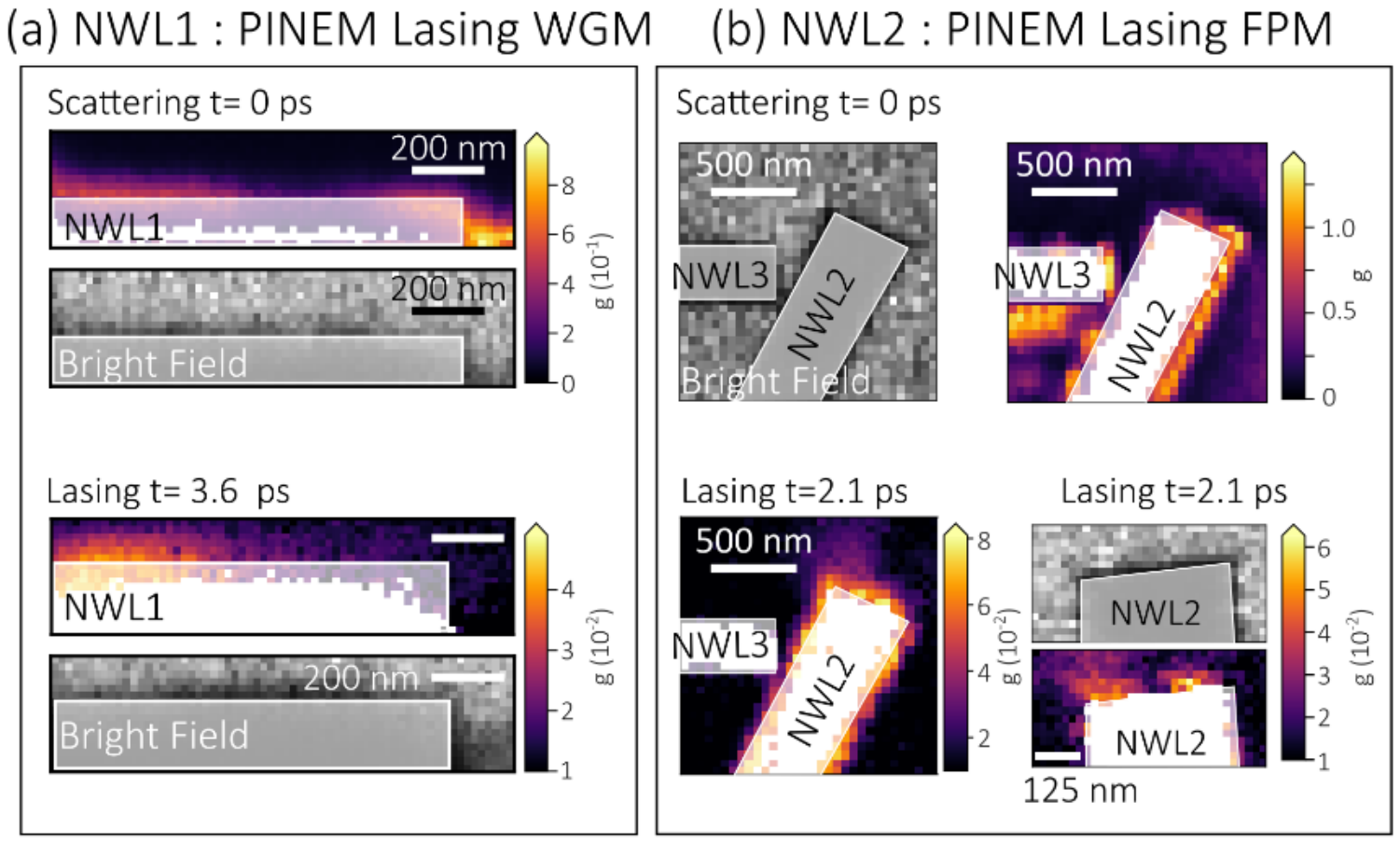

**Fig. S15** : PINEM Map of NWL1 and NWL2 at Δt = 0 ps. The PINEM at $\Delta t > t_{onset}$ is shown as a reference

## XI. Delay Scan without Electron chirp above the lasing threshold

We present here some data taken without chirp on another NWL. The laser power used for photoemission was significantly reduced in order to reduce the number of electrons produced at the tip to less than one electron per pulse, hence avoiding coulomb repulsion and an increase in the energy spread.

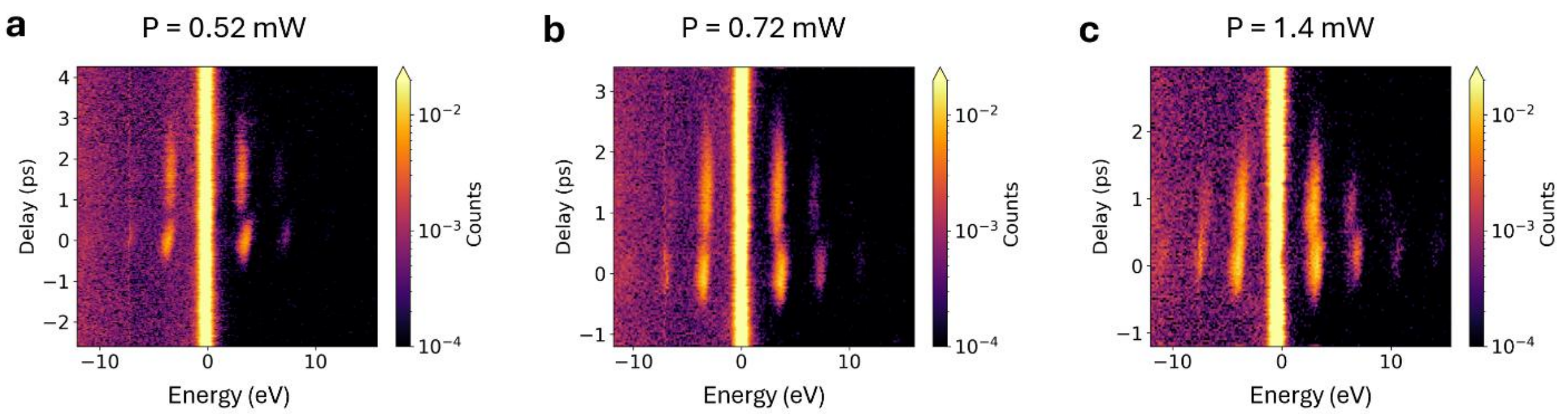

**Fig. S16**: **Delay scan of NWL5**. Delay scan where the photoemission condition was set to reduce the chirp to minimum. Delay scan at the same position for different pump power: P = 0.52 mW (a), P = 0.72 mW (b), P = 1.4 mW (c).

**Fig. S16** shows delay scans for a minimum chirp of the electron beam, we can see no tilt of the first side band corresponding to the scattering of the laser. The second set also does not show widening with delay, unlike the data presented in **Fig. 2** of the main article.